\documentstyle[12pt]{article}

\def\MeV{\,{\rm MeV}}
\def\sec{\,{\rm sec}}

\def\yr{\,{\rm yr}}
\def\rcm{\,{\rm cm}}

\def\Mpc{\,{\rm Mpc}}

\def\eV{{\,\rm eV}}

\def\erg{{\,\rm erg}}
\def\cmm2{{\,\rm cm^{-2}}}
\def\cm2{{\,{\rm cm}^2}}
\def\cmm3{{\,{\rm cm}^{-3}}}
\def\gcmm3{{\,{\rm g\,cm^{-3}}}}
\def\kms{\,{\rm km\,s^{-1}}}

\def\mpl{{m_{\rm Pl}}}

\def\la{\mathrel{\mathpalette\fun <}}

\def\fun#1#2{\lower3.6pt\vbox{\baselineskip0pt\lineskip.9pt
  \ialign{$\mathsurround=0pt#1\hfil##\hfil$\crcr#2\crcr\sim\crcr}}}

\begin{document}
\pagestyle{empty}
\begin{center}
\rightline{FERMILAB--Pub--93/134-A}
\rightline{to appear in {\it Science}}

\vspace{.2in}
{\Large \bf WHY IS THE TEMPERATURE OF THE UNIVERSE 2.726\,K?} \\

\vspace{.2in}
Michael S. Turner\\

{\it Departments of Physics and of Astronomy \& Astrophysics\\
Enrico Fermi Institute, The University of Chicago, Chicago, IL~~60637-1433}\\

\vspace{0.1in}

{\it Theoretical Astrophysics\\
Fermi National Accelerator Laboratory, Batavia, IL~~60510-0500}\\

\end{center}

\vspace{.3in}

\centerline{\bf ABSTRACT}
\noindent NASA's Cosmic Background Explorer (COBE) Satellite has recently
made the most accurate measurement of the temperature
of the Universe determining it to be $2.726\pm 0.01\,$K.
In trying to understand why the temperature has this value,
one is led to discover the most fundamental features
of the Universe---an early, radiation-dominated epoch,
enormous entropy per nucleon, synthesis of the
light elements around three minutes after the bang,
and a small excess of matter over
antimatter---as well as some of the most pressing issues in cosmology today,
the development of structure in the Universe
and the identification of the nature of the ubiquitous dark matter.

\newpage
\pagestyle{plain}
\setcounter{page}{1}
\newpage

\section{The Cosmic Background Radiation}

The existence of the cosmic background radiation (CBR) is one
of the cornerstones of the standard cosmology, or hot big-bang
model \cite{std}.  Indeed, its very existence provides the evidence that
the Universe began from a hot state \cite{hot}.
The temperature of the cosmic background radiation
has recently been measured to unprecedented precision
by the Far InfraRed Absolute Spectrophotometer (FIRAS)
instrument on NASA's Cosmic Background
Explorer (COBE) Satellite \cite{COBETEMP}:
\begin{equation}
T_0 = 2.726\,{\rm K}\pm 0.01\,{\rm K} ;
\end{equation}
the FIRAS results are shown in Fig.~1.

Measurements of the CBR temperature, made over the period
of almost thirty years since its discovery by Penzias
and Wilson \cite{PW}, now span almost three and half
decades in wavelength, from about $0.04\rcm$ to $70\rcm$,
and are all consistent with the COBE temperature.
Deviations from a perfect black-body spectrum
are less than 0.03\% over the wavelengths
probed by COBE, $0.05\rcm - 0.5\rcm$ \cite{COBETEMP}.
The CBR is probably the most well studied
and best black body known; indeed, the COBE
collaboration plans to use their data
to test the form of the Planck Law itself \cite{Mather}.

With a number density of $411\cmm3$ the photons in the
CBR by a wide margin account for most of the (known) particles in the
Universe, outnumbering atoms
by a factor of around a billion.  The surface of last scattering
for the CBR is the Universe itself at an age of a few 100,000 years
(see Fig.~2), and thus the CBR provides a fossil record of the
infant Universe.
As such its every property has been studied---spectrum,
polarization, and spatial isotropy---revealing
important information about the evolution of the Universe \cite{sz}.
As I will discuss, just trying to answer the simple question,
why is the temperature of the CBR $2.726\,$K?,
reveals the most fundamental features of the Universe as well as
several pressing problems in cosmology.

To begin, it is imprecise to say that the Universe
has a temperature, as it is not in thermal equilibrium today.
Earlier than a few 100,000 years the matter was ionized
and a state of thermal equilibrium existed; at about
this time the temperature was about $3000\,$K
and the equilibrium ionization fraction of matter became very small.
The Universe is said to have ``recombined;'' since neutral
matter is transparent to the radiation, the CBR photons
we detect today last scattered a few 100,000 years after
the bang.  After last scattering, the expansion
simply red shifted the energy of CBR photons and
diluted their number density, and, because of a remarkable
feature of the expansion, a Planck distribution
was maintained with a temperature that decreased in proportion to the
size of the Universe.  For this reason,
the Universe today is filled with thermal radiation
of temperature $2.726\,$K despite the fact that the Universe
is no longer in thermal equilibrium.

Since the temperature of the Universe is decreasing---and
has been for some 15 billion years or so---the original question
must be rephrased:  Why did the temperature
of the Universe reach about $3\,$K at an age
of about 15 billion years old?  (Several independent
measures of the age, based upon the evolution of stars in
the oldest globular clusters, the cooling of the oldest
white dwarfs in the Galaxy, and the dating of certain radioactive
isotopes, indicate that the Universe is between
12 and 18 billion years old \cite{age}.)

According to Einstein's equations, the present age of the Universe---that is,
time since the bang---is related to the present energy density:
\begin{equation}\label{eq:age}
t_0 = {1\over \sqrt{6\pi G \rho_0}};
\end{equation}
where $G$ is Newton's gravitational constant and
for simplicity I have assumed that the Universe is spatially flat
($\Omega_0=1$).  The quantity
$\Omega_0 =\rho_0 /\rho_{\rm crit}$ is the ratio
of the mean density to the critical, or closure, density; $\rho_{\rm crit}
= 3H_0^2/8\pi G \simeq 1.88\times 10^{-29}\,
(H_0/100\kms \Mpc^{-1})^2\,\gcmm3$ and $H_0 = 40\kms \Mpc^{-1}
-100\kms \Mpc^{-1}$ is Hubble's constant, whose value is still
only known to within a factor of two.  ``Low-density''
universes, $\Omega_0<1$, are negatively
curved and expand forever, while ``high-density''
universes, $\Omega_0>1$, are positively curved and eventually recollapse.
The ``critical'' universe, $\Omega_0=1$, is spatially flat and
also expands forever.
In the general case, $t_0 =\sqrt{3\Omega_0 f^2(\Omega_0 )/8\pi G\rho_0}$,
where the function $f(\Omega_0)$ varies between 1
and $2/3$ for $\Omega_0$ between 0 and 1.

We know at least one component of the energy density today:
the CBR black-body radiation itself, which contributes an energy density
\begin{equation}
\rho_{\rm CBR} = {\pi^2 k_B^4 T_0^4\over 15 \hbar^3 c^3}
\simeq 4.18\times 10^{-13}\erg \cmm3 ,
\end{equation}
where $\hbar = 1.05\times 10^{-27}\erg\sec$
is Planck's constant divided by $2\pi$,
$k_B = 1.38\times 10^{-16}\erg\,{\rm K}^{-1}$
is Boltzmann's constant, $c =3.00\times 10^{10}\rcm\sec^{-1}$
is the speed of light,
and $\pi^2 k_B^4/15\hbar^3 c^3=7.56\times 10^{-15}
\erg\rcm^{-2}\sec^{-1}\,{\rm K}^{-4}$ is four times the Stefan-Boltzmann
radiation constant divided by $c$.  If the CBR were the only
contribution to the energy density, Eq.~(\ref{eq:age})
would imply an age of about 1300 billion years, a factor
of about 100 too large.  Put another way, for the
age to be consistent with the energy density in the
CBR alone, the temperature would have to be closer to $30\,$K.

\section{Matter in the Universe}

By asking a simple question we have learned that
the CBR black-body radiation must make a minor contribution to the energy
density today, $\rho_{\rm CBR}\sim 10^{-4}\rho_0$.
What then accounts for the bulk of the present energy density?
It could exist in other
thermal backgrounds of relativistic particles; however,
that would require the existence of several
thousand additional massless particle species---and we know
of at most three, the electron, muon, and tau
neutrinos, which together contribute a energy density
comparable to that of the CBR (provided all three neutrino
species are massless).

It is almost certain that the bulk of the energy density exists in the form of
nonrelativistic matter \cite{krauss}.  Taking the
age of the Universe to be 15 billion years, Eq.~(\ref{eq:age})
implies a matter density of about $3\times 10^{-30}{\rm g}
\cmm3$ (energy density of about $3\times 10^{-9}\erg
\cmm3$).  Today the energy density
in matter is more than ten thousand times greater than that
in the CBR, but that was not always the case.  As
the Universe expands the matter density decreases as $R^{-3}$,
by the factor by which the volume increases; $R(t)$ is
the cosmic-scale factor which describes the linear
expansion of the Universe.  The energy density in radiation
decreases faster, as $R^{-4}$, because the energy
of each photon is also ``red shifted'' by the expansion,
accounting for the additional factor of $R^{-1}$.
Owing to the different scalings of the matter
and radiation energy densities, when the Universe
was about $10^{-4}$ of its present size and a few thousand
years old, the two energy densities
were equal.  Earlier than this the energy density in
radiation exceeded that in matter, and the Universe is
said to be ``radiation dominated.''

Early on matter was a trace
constituent in a Universe dominated by a hot plasma of
thermal particles; at the earliest times, $t\ll 10^{-5}\sec$,
the hot plasma was a soup of the fundamental particles, quarks, leptons, and
gauge bosons (the photon, $W^\pm$ and $Z^0$, and gluons,
the carriers of the forces).  This is an extremely important
feature of the Universe and has profound implication for
the study of its earliest history.
Among other things it means that the formation of
structure in the Universe---galaxies, clusters of galaxies, voids,
superclusters, and so on---through the gravitational amplification
of small inhomogeneities in the matter density only began
a few thousand years after the bang \cite{sf}.  This is because
during the radiation-dominated phase the self-gravitational
attraction of the matter was no match for the rapid expansion
driven by the enormous energy density in radiation, and density
perturbations could not grow (see Fig.~3).

A year ago, another instrument on the COBE satellite,
the Differential Microwave Radiometer (DMR), detected tiny differences
in the CBR temperature measured in different directions:  on average
about a part in $10^5$ (or $30\mu\,$K) between directions separated by
$10^\circ$; see Fig.~4 \cite{DMR}.  Inhomogeneities
in the matter density give rise to temperature variations of a similar
size, and this COBE discovery provided the first
evidence for the existence of the primeval density
fluctuations that seeded all the structure in the Universe.
Moreover, since density fluctuations grow in proportion
to the cosmic-scale factor and the level of inhomogeneity
($\delta \rho /\rho$) exceeds unity today,
the amplitude of the primeval fluctuations needed to seed
the observed structure is set roughly by
the size of the scale factor when the matter and radiation energy
densities were equal---about $10^{-5}$ or so---a number which
is determined by the present ratio of the energy density
in radiation to that in matter.  In a very real sense, the CBR temperature
set the amplitude of temperature fluctuations that were expected!

The extreme uniformity of the temperature of the CBR across
the sky, to better than a part in $10^4$ on angular scales
from arcminutes to $180^\circ$ \cite{smallscale}, reveals
an important property of the Universe---its smoothness,
or isotropy and homogeneity---and raises another
question---why is it so smooth?  Though the Universe
was very small at early times, its rapid expansion limited the
distance over which even photons could travel.  At the
epoch of last scattering this distance corresponds to an angle
of only about $1^\circ$ on the sky; this fact precludes any
causal physical process from explaining the temperature uniformity, and
hence the smoothness of the Universe, on angular
scales greater than this.  Further, it raises
the same question about the origin
of the primeval density inhomogeneities; they too could not
have been created on such large distances by causal processes
operating at early times.

The smoothness and the primeval inhomogeneity needed
to seed structure could have existed since the beginning.
However, Guth showed that both can be explained by a very rapid
period of expansion---called cosmic inflation---that may
have taken place about $10^{-34}\sec$ after the bang \cite{inflation}.
This rapid expansion is driven by the false-vacuum
energy (particle physics analogue of latent heat)
associated with a first-order phase transition.
The basic idea is that a tiny patch
of the Universe, which could have been made
smooth early on, grew exponentially to a size that would
encompass all that we can see today and well beyond.
The enormous growth of the scale factor also allows quantum
mechanical fluctuations arising during inflation
on small length scales to become density perturbations on length
scales large enough to account for the primeval density inhomogeneities needed
to seed the structure seen today \cite{bst}.  (The COBE DMR
results are consistent with the temperature variations predicted in
inflationary models, as are two other models for the origin
of the density fluctuations.)
In addition, the tremendous growth
in the size of the Universe---by a factor greater than that
by which the Universe has grown since---also leads
to a Universe that, regardless of its initial curvature,
today still appears flat, making $\Omega_0=1$ a ``prediction'' of inflation.

\section{The Nucleon-to-Photon Ratio}

Assuming that the present mass density exists in the
form of ordinary matter, atoms made of nucleons---neutrons
and protons---and electrons, a present nucleon density
of about $2\times 10^{-6}\cmm3$ is implied.
{}From this we can form the dimensionless ratio of
the nucleon-number density to the photon-number density:
\begin{equation}
\eta \equiv {n_N \over n_\gamma} \sim 5\times 10^{-9}.
\end{equation}
This ratio indicates that CBR photons outnumber nucleons by a
factor of around a billion.  The inverse of $\eta$,
the ratio of photons to nucleons, measures
the entropy in radiation per nucleon (in units of $k_B$).
The radiative entropy per nucleon in a star like our
sun is only about $10^{-2}$; even in the highest entropy
environment known, the center of a newly born neutron
star, the entropy per nucleon is only a few.
The Universe has an extremely high entropy, so high
that it is very difficult to imagine that any physical process
could have produced the CBR or added significantly
to it.  Further, because the CBR spectrum is so accurately
Planckian, there are severe restrictions on any process
that produces photons, e.g., radiation from an early generation of stars
or the decay of relic neutrinos (if they are unstable).
In any case, the entropy per nucleon {\it seems} to be an initial condition,
rather than a quantity that is easily explained.

The nucleon to photon ratio $\eta$ also quantifies
the net excess of nucleons over antinucleons, or net
baryon number, per photon.  The net baryon
number per photon is equal to $\eta$ because
there is no significant amount of antimatter
in the Universe today (i.e., $n_{\bar N}\ll n_N$):
\begin{equation}
{n_B\over n_\gamma} \equiv {n_N-n_{\bar N} \over n_\gamma}
\approx {n_N\over n_\gamma}.
\end{equation}
Baryon number, like charge, is known empirically to be conserved to
a high degree of precision.  (The longevity of the proton, lifetime greater
than $10^{32}\yr$, attests to this; were baryon number
not conserved, the proton would be expected to decay in
a fraction of a second.)  Conservation (or even approximate conservation)
of baryon number and the value of $\eta$ imply
that earlier than about $10^{-5}\sec$,
when it was hot enough for matter and antimatter to
be freely created, there was approximately one more
baryon than antibaryon for every billion or so of both.
Looking at it the other way around, in the absence of this tiny
excess, all the baryons and antibaryons would have annihilated
as the Universe cooled leaving essentially no matter or antimatter today.

Though the details have not been worked out,
many believe that this excess of matter over
antimatter so crucial to the existence of matter
today, evolved due to particle interactions in the
very early Universe ($\la 10^{-12}\sec$) that neither respect
the symmetry between matter and antimatter nor
the conservation of baryon number \cite{baryo}.
(Violation of the conservation of baryon number is
an almost universal prediction of theories
that attempt to unify the forces of Nature,
and also arises in the standard model of particle physics
due to subtle quantum mechanical effects.  The symmetry
between matter and antimatter is observed to be violated
by a small amount in the decays of the $K^0,{\bar K}^0$ mesons.)
Explaining the small net baryon number, quantified by
$\eta$, appears to be much
more promising than trying to explain the large entropy,
quantified by $\eta^{-1}$.

The high entropy plays a crucial role in determining the
chemical composition of the Universe.  Were
the entropy per nucleon even
a thousand times smaller, nuclear reactions taking
place when the Universe was only a fraction of a second old and
the energy equivalent of the temperature $k_BT$ was
${\rm few}\,\MeV$ would
have quickly processed all the nucleons into tightly
bound nuclei such as carbon, oxygen and on up to iron.  Instead,
most of the nucleons remain
in the form of protons with only the lightest isotopes,
D, $^3$He, $^4$He, and $^7$Li, being produced.  (It is
generally believed that the other elements were
produced in stars or spallation reactions in the interstellar
medium.)  The lack of significant nucleosynthesis beyond the light
elements traces directly to
the high entropy:  The enormous number of high-energy photons
per nucleon delays the onset of nucleosynthesis until
a temperature of order $k_BT\sim 0.1\MeV$ because earlier
photons rapidly dissociated nuclei as they formed;
when nucleosynthesis did begin coulomb repulsion
between light nuclei prevented their
fusion into the heavier, more tightly bound nuclei.
(This fact was appreciated before the discovery of
the CBR and led Gamow and others to predict the existence
of a relic radiation with about the correct temperature \cite{gamow}.)

The predictions of primordial nucleosynthesis agree with
the inferred primordial abundances of the light elements
provided that the nucleon-to-photon ratio lies in the interval
\begin{equation}
3\times 10^{-10} \la \eta \la 4\times 10^{-10}.
\end{equation}
The very existence of a ``concordance interval''
is an important test of the standard cosmology,
and as a bonus it provides the most accurate determination of
the nucleon-to-photon ratio \cite{walkeretal}.
The predictions of big bang nucleosynthesis and
the observed abundances
of the light elements are shown in Fig.~5.

The success of the theory of primordial nucleosynthesis
not only provides the earliest test of the big-bang
model, but it also leads to a startling suggestion:
that most of the matter in the Universe is something
other than nucleons.
{}From primordial nucleosynthesis and the temperature
of the CBR the mass density contributed by nucleons
can be computed:
\begin{equation}
\rho_N = m_N \eta n_\gamma \simeq 2.7 \times 10^{-31}\,
{\rm g}\cmm3 ,
\end{equation}
where $m_N\simeq 1.7\times 10^{-24}\,{\rm g}$
is the mass of a nucleon, $n_\gamma = 2\zeta (3)
k_B^3T_0^3 /\pi^2 \hbar^3 c^3 =411\rcm^{-3}$ is the number density of
photons, and $\zeta (3) = 1.20206\cdots$.
This is lower than the earlier estimate of the total mass
density derived from the age of the Universe, though to be sure,
we made certain assumptions at the time.  In any case, the
small mass density in nucleons leads one to ask whether the mass density
of the Universe is greater than that contributed by ordinary matter alone.

\section{Dark Matter in the Universe}

Let me very briefly review what we know about
the mass density of the Universe \cite{mass}. Based upon
the above determination of the density of ordinary matter
and our imperfect knowledge of the Hubble constant,
it follows that ordinary matter contributes between 1\% and 10\% of the
critical density (the larger value for the lower value
of the Hubble constant).  From astronomical observations
we know:  (i) luminous matter, in the form of stars,
contributes less than 1\% of critical density; (ii) other
observations that measure the amount of mass through its
gravitational effects, e.g., the motion of stars in spiral
galaxies \cite{rubin}, the motions of galaxies in clusters,
and so on, indicate that the total amount of mass is {\it at
least} 10\%-20\% of the critical density \cite{iau}; (iii) our motion with
respect to the CBR suggests that the density is near
critical; and (iv) no definitive
measurement of the total amount of matter has yet been made(!).

The third point deserves further discussion; the CBR is hotter
in the general direction of the constellations Hydra and
Centaurus, by about 3\,mK, and cooler in
the opposite direction by the same amount \cite{dipole}; see Fig.~6.
The simplest---and now standard---interpretation is
that our galaxy is moving with respect to the ``cosmic rest frame''
at a speed of about $620\kms$.  (It is interesting
to note that COBE detected a much smaller
yearly modulation of the same kind arising due to Earth's motion around the
sun at $30\kms$; this should convince any remaining
``geocentrists'' that the Earth does indeed move!)
The motion of the Milky Way arises due to the gravitational
tugs exerted on it by the thousands of galaxies within a hundred Mpc or so.
Because the distribution of
galaxies is not precisely homogeneous, the sum of these tugs
does not cancel, but results in a net force in the direction
of Hydra-Centaurus.  Since the gravitational force on the Milky Way
due to another galaxy is proportional to that galaxy's mass,
an estimate for the mass in this volume---and for the
average mass density---can be made by
relating our velocity to the observed distribution
of galaxies in this volume.  This technique samples the
largest volume of space of any method yet, and
indicates a value for $\Omega_0$ that is close to unity \cite{iras}.

Though our knowledge of the mass density of the
Universe is still incomplete, we can already conclude that:  (i) most
of the matter in the Universe is dark, i.e., does not
emit or absorb radiation of any wavelength; (ii) if the mass density
of the Universe is at the lower limit of current estimates
and if the density of ordinary matter is at its upper limit,
then ordinary matter {\it could} account for all the mass
with $\Omega_0$ being around 0.1; (iii) on the other hand, if the mass density
is significantly greater than
10\% of the critical density, then the dark matter must be
something other than ordinary matter.
This possibility is favored by many cosmologists, mainly the
theorists, as theoretical considerations, including cosmic
inflation and theories of structure formation,
argue strongly for the critical Universe ($\Omega_0=1$).
I hasten to add the observational situation is far from settled,
and many, if not most, astronomers would say that the case
for $\Omega_0 =0.1$ is the more compelling one at present.

It is interesting to note the crucial role played by the CBR temperature
in reaching these conclusions.
The outcome of primordial nucleosynthesis depends only
upon the nucleon-to-photon ratio.  Therefore
the primordial abundances of the
light elements serve to determine $\eta$ rather than the nucleon
mass density itself.  To determine nucleon mass density the photon-number
density---and hence CBR temperature---must be known.
Were the CBR temperature a factor of three or so higher,
the mass density contributed by ordinary matter
would be close to the critical density.

If most of the mass in the Universe is not
ordinary matter, what is it?  The most promising idea
is that it exists in the form of elementary particles left
over from the early, fiery moments of the
Universe \cite{relic}.  In this case, another dimensionless ratio can be
formed,
the ratio of the number density of ``exotic particles'' to CBR photons,
\begin{equation}
\eta_X \equiv {n_X\over n_\gamma} \simeq
7\times 10^{-9} \left( {m_N\over m_X} \right),
\end{equation}
where $m_X$ is the mass of the exotic and
for simplicity I have assumed that exotic particles contribute
critical density and a Hubble constant of $50\kms\Mpc^{-1}$.

As it turns out, there are a handful of interesting
candidates for the dark matter.
They include a massive neutrino; the neutralino;
and the axion.  All three possibilities are motivated
by particle-physics considerations first with
their important cosmological
consequence as a bonus---and perhaps even a hint that
the particle dark-matter hypothesis is on the right track.

How do these particles arise as relics of the big bang?
In the early Universe thermodynamics dictated a kind of particle
democracy, with all species being roughly equally abundant.
As the Universe cooled pair creation of massive particles
became energetically forbidden, and massive particle species disappeared
through particle-antiparticle annihilations.   If a particle
species is stable, it can have a significant relic
abundance because in the expanding Universe annihilations
eventually cease as particles and antiparticles become too
sparse to encounter one another and annihilate.  The relic abundance
depends upon the potency of annihilations, quantified
as the annihilation cross section, $\sigma_{\rm ann}$,
which has units of area.

In the case of neutrinos, annihilations became ineffective
before they could start significantly reducing the
neutrino abundance relative to photons, and so
$\eta_X$ is expected to be around one (more precisely
3/11).   Thus the contribution of neutrinos to the
mass density is dictated by their mass:  They contribute
critical density for a mass of about $2.5\times 10^{-8}
m_N$, or a mass energy of about twenty electronvolts
(eV).  Such a mass is in the ballpark predicted for
neutrino masses by many unified theories of particle
interactions \cite{seesaw}.  While experimental evidence rules out
a mass this large for the electron neutrino, it is still
possible that either the muon or tau neutrino has such a mass.

The neutralino is a particle that is predicted to exist
in supersymmetric extensions of the standard model of particle
physics \cite{susy}; predictions for its mass are rather uncertain,
ranging from ten to thousand times that of the
nucleon.  (Supersymmetry dictates a spin one-half partner
for every integer spin particle, and vice versa;
in the simplest supersymmetric
models the neutralino is the spin one-half partner of the
photon.)  In the case of the neutralino, annihilations
significantly decrease the number of neutralinos from their
early abundance of one per photon.  Their relic abundance
is inversely proportional to their annihilation cross section,
very roughly
\begin{equation}
\eta_X \sim {(\hbar /c)^2 \over m_X \mpl \sigma_{\rm ann}},
\end{equation}
where $\mpl = \sqrt{\hbar c/G} \simeq 2.2\times 10^{-5}\,$g
is the Planck mass.  Note that the relic abundance
depends inversely upon the neutralino mass, so that
it cancels out when computing the relic mass
density of neutralinos.  Remarkably, the condition that the neutralino
contribute critical density becomes a condition on
its annihilation cross section alone,
\begin{equation}
\sigma_{\rm ann} \sim {10^{-2} \hbar^2 \over k_BT_0\,\mpl}
\sim 10^{-36}\rcm^2 .
\end{equation}
The cross section required is of the order of magnitude
of a weak-interaction cross section, which is the general size
expected for the neutralino annihilation cross section.

The axion is a particle whose existence traces to trying
to solve a nagging problem of the standard model of particle
physics, the strong-$CP$ problem.  Subtle quantum mechanical effects
associated with Quantum-Chromo-Dynamics (QCD), the theory of the
strong interactions that bind quarks together, result in a predicted
value for the electric-dipole moment of the neutron
that is nine orders of magnitude larger than the current
experimental upper limit.  In 1977 Peccei and Quinn proposed an
elegant solution:  the introduction of a new symmetry (now
referred to as PQ symmetry) that solves the problem and leads to
the prediction of a new particle, the axion \cite{PQ}.  The axion
interacts more feebly than neutrinos do, which explains
why its existence has yet to be verified or falsified,
and, for the same reason it would not have been
produced in the thermal plasma during the earliest moments.

Relic axions arise in a different and rather unusual way.
Because the axion interacts so weakly, the value of the axion field
is left undetermined at early times, taking on whatever random value
it had at the beginning; eventually, at about $10^{-5}\sec$,
due to QCD effects, the axion field begins to relax to its equilibrium
value.  In so doing, it overshoots that value and is left
oscillating.  These cosmic harmonic oscillations correspond
to an extremely high density of very low momentum axions
that should still be with us today.  If the rest mass energy of the
axion is around $10^{-5}$ electronvolts relic axions provide
closure density \cite{axion1}.  Theoretical considerations
do little to pin down the mass of the axion;  however, a host
of laboratory experiments and astrophysical/cosmological arguments
have narrowed the allowed window for its mass to $10^{-6}\eV$
to $10^{-3}\eV$, roughly the range where it would contribute
close to the critical density \cite{axion2}.

All three particle candidates for the dark
matter are sufficiently attractive that experimental
efforts are underway to test their candidacies \cite{dmsearch}; in the
case of the axion and neutralino, the experiments involve
actually detecting the particles that
comprise the dark halo of our own galaxy \cite{halodm}.
For the neutrino, direct laboratory measurements restrict the
electron-neutrino mass to be less than about $8\eV$, too
small to account for the critical density.  Direct measurements
of the muon and tau neutrino masses are far more difficult and
cannot come close to probing a mass as small as
$20\eV$; indirect experiments, such as neutrino oscillation
experiments and solar neutrino observations, can provide
some information, but thus far nothing conclusive \cite{numass}.

\section{Development of Structure in the Universe}

One of the most pressing questions in cosmology concerns
the details of how the abundance of structure seen in the
Universe today came to be.
If the bulk of the matter in the Universe exists in the form
of particle relics from the big bang there are profound
implications for how structure formed.
First, the process can begin earlier, as soon as the
Universe becomes matter dominated, about 1000 years
after the bang; if there is only ordinary
matter the growth of the primeval density perturbations
cannot begin until matter and radiation
decouple, a few 100,000 years after the bang, when matter
is freed from the drag of the radiation.  Because density
inhomogeneities can start growing sooner, their initial
amplitude can be smaller, leading to smaller predicted variations
in the CBR temperature.

The COBE DMR result is consistent with this smaller prediction,
but by no means confirms the existence of exotic dark matter.
One of the three viable scenarios of structure
formation involves ordinary matter only.  This minimalist
picture, proposed by Peebles \cite{PIB}, postulates a Universe
with baryonic matter only, the dark
matter existing in the form of ``dark'' stars (low-mass stars or
the remnants of high-mass stars---neutron stars or black holes).
The density fluctuations
arise from local fluctuations in the number of baryons (of unknown
origin) and the spectrum is adjusted to both explain the observed
structure and to be consistent with the level of CBR anisotropy.
The weak point of this model is that $\Omega_0$ must be at least
0.2 in order to form the observed structure,
which violates the nucleosynthesis bound since all the matter
is baryonic.

There are two broad classes of models for structure formation
with particle dark matter:  hot dark matter, where the dark
matter exists in the form of neutrinos, and cold dark matter, where
it exists in the form of neutralinos
or axions.  In the case of hot dark matter the primeval density fluctuations
on small length scales are erased by the streaming of fast
moving neutrinos from regions of higher density into those of
lower density, and the structures that form first are very
large---superclusters---and smaller structures---galaxies and
so on---must be formed by fragmentation.  This so-called ``top-down''
scenario is disfavored as structures as large as superclusters
are just forming today, making it difficult to explain the
existence of distant galaxies that must have formed long ago \cite{hdm}.

The erasing of fluctuations on small length scales does not
occur with cold dark matter because the dark-matter particles
move very slowly---neutralinos because they are so heavy and
axions because they were born with very low momentum.  With
cold dark matter structure develops ``bottom-up,'' from
galaxies to clusters of galaxies to superclusters.
Cold dark matter seems to work much better,
though not perfectly \cite{cdm}.  It has been
suggested that the cold dark matter scenario
could be improved by ``mixing'' in a small
amount of hot dark matter, in the form of neutrinos of
mass $7\eV-10\eV$, referred to as mixed dark matter \cite{MDM}.

To complete the description of a scenario for structure
formation the origin of the primeval fluctuations must be specified.
One possibility involves
quantum fluctuations arising during inflation, as discussed
earlier.  This leads to the fairly successful (in this author's
opinion) and very well studied ``cold dark matter'' scenario.
Another possibility is that the primeval fluctuations involve
topological defects---monopoles, string, or texture---that
act as gravitational seeds and were produced in a cosmological phase
transition that occurred about
$10^{-36}\sec$ after the bang.  These scenarios are less
well developed, but look promising \cite{top}.

At present there are three viable pictures of structure
formation, two early Universe scenarios---inflation produced
density fluctuations plus cold
dark matter and topological defects plus cold (or possibly hot)
dark matter---and the minimalist scenario involving only ordinary matter.
Further study of the the tiny variations in the CBR temperature
on angular scales of order $1^\circ$ should soon help to whittle down the list.

\section{Conclusion}

The cosmic background radiation is arguably
the most important cosmological relic yet discovered, and much has
and will be learned from its study.  The CBR is so fundamental to the
standard cosmology that just trying to understand why
its temperature is $2.726\,$K today leads one to discover
the most fundamental features of the Universe as
well as some of the most pressing cosmological
problems---the origin of structure and the nature of
the dark matter.  In the end, we have no firm explanation
as to why the Universe even has a temperature;
that is, where the fiery radiation came from.
According to the inflationary scenario its existence traces to the
decay of the false-vacuum energy.  However, its
explanation, like that of the expansion itself,
may well involve physics yet to be understood.

\bigskip\bigskip\bigskip
This work was supported in part by the Department of Energy
(at Chicago) and by the NASA through grant NAGW-2381 (at Fermilab).

\vfill\eject
\section{Figure Captions}
\bigskip

\noindent {\bf Figure 1:}  The COBE FIRAS measurements of the
CBR spectrum and the spectrum of a 2.726\,K black body.  Note
the COBE one-sigma error flags have been enlarged by a
factor of 100.

\medskip
\noindent {\bf Figure 2:}  A schematic diagram illustrating the
last-scattering surface.  Also shown is the angle subtended
on the sky by the photon travel distance from the bang until
the time of last scattering.

\medskip
\noindent {\bf Figure 3:}  The growth of primeval density perturbations
and the ratio of energy density in the CBR to that in matter as
a function of cosmic-scale factor.  With ordinary matter only,
perturbations begin growing when matter and radiation decouple;
with particle dark matter
perturbations begin growing much earlier, as soon as the Universe becomes
matter dominated, and thus smaller primeval density inhomogeneities
are required.

\medskip
\noindent {\bf Figure 4:}  The COBE DMR measurements of the tiny
variation in the temperature between points on the sky separated
by angle $\theta$.  (Note, the anisotropy due to the motion
of the Milky Way has been removed.)

\medskip
\noindent {\bf Figure 5:}  The predictions of primordial nucleosynthesis
and the inferred primordial abundances of D, $^3$He, $^4$He, and
$^7$Li.  The concordance interval is shaded.

\medskip
\noindent {\bf Figure 6:}  The COBE DMR temperature map of the sky.
The variation in the temperature on the sky is represented
on a color scale (pink is hot, blue
is cold); the dipole anisotropy is clearly seen.


\begin{thebibliography}  {std}

\bibitem{std} For a textbook treatment of the standard
cosmology see e.g., S.~Weinberg, {\it Gravitation and
Cosmology} (Wiley, NY, 1972); P.J.E.~Peebles, {\it Principles
of Physical Cosmology} (Princeton University Press, Princeton, 1993);
or E.W.~Kolb and M.S.~Turner, {\it The Early
Universe} (Addison-Wesley, Redwood City, CA, 1990).
At a more readable level, though slightly out of date,
S.~Weinberg, {\it The First Three Minutes} (Basic Books, NY, 1979).

\bibitem{hot}  R.H.~Dicke, P.J.E.~Peebles, P.G.~Roll,
and D.T.~Wilkinson, {\it Astrophys. J.} {\bf 142}, 414 (1965).

\bibitem{COBETEMP} J.~Mather et al., {\it Astrophys. J.}
{\bf 354}, L37 (1990); {\it ibid}, in press (1993).  Another
very accurate measurement of the CBR temperature was
made from a rocket-borne instrument by H.~Gush, M.~Halpern, and E.H.~Whishnow
{\it Phys. Rev. Lett.} {\bf 65}, 537 (1990).  Reviews of
earlier measurements of the CBR spectrum include
R.~Weiss, {\it Ann. Rev. Astron. Astrophys.} {\bf 18}, 489 (1980),
and G.~Smoot et al., {\it Astrophys. J.} {\bf 331}, 653 (1988).

\bibitem{PW} A.A.~Penzias and R.W.~Wilson, {\it Astrophys. J.}
{\bf 142}, 419 (1965).

\bibitem{Mather} J.~Mather, private communication (1993).

\bibitem{sz} See e.g., R.A.~Sunyaev and Ya.B.~Zel'dovich,
{\it Ann. Rev. Astron. Astrophys.} {\bf 18}, 537 (1980).

\bibitem{age} See e.g., J.J.~Cowan, F.K.~Thielemann,
and J.W.~Truran, {\it Ann. Rev. Astron. Astrophys.} {\bf
29}, 447 (1991), or W.A.~Fowler, {\it Q. Jl. R. astr. Soc.}
{\bf 28}, 87 (1987).

\bibitem{krauss} The possibility that the Universe today
is radiation-dominated is discussed by
M.S.~Turner, G.~Steigman, and L.~Krauss,
{\it Phys. Rev. Lett.} {\bf 52}, 2090 (1984).

\bibitem{sf}  See e.g., P.J.E.~Peebles, {\it The Large-scale
Structure of the Universe} (Princeton University Press, Princeton, 1980);
G.~Efstathiou, in {\it The Physics of the Early Universe}, eds.~J.A.~Peacock,
A.F.~Heavens, and A.T.~Davies (Adam-Higler, Bristol, 1990).

\bibitem{DMR} G.F.~Smoot et al., {\it Astrophys. J.} {\bf 396},
L1 (1992).

\bibitem{smallscale} See e.g., A.C.S.~Readhead and C.R.~Lawrence,
{\it Ann. Rev. Astron. Astrophys.} {\bf 30}, 653 (1992).

\bibitem{inflation} A.H.~Guth, {\it Phys. Rev. D} {\bf 23},
347 (1981); A.D.~Linde, {\it Phys. Lett. B} {\bf 108},
389 (1982); A.~Albrecht and P.J.~Steinhardt, {\it Phys. Rev. Lett.}
{\bf 48}, 1220 (1982).

\bibitem{bst} J.M.~Bardeen, P.J.~Steinhardt, and M.S.~Turner,
{\it Phys. Rev. D} {\bf 28}, 679 (1983);
A.H.~Guth and S.-Y.~Pi, {\it Phys. Rev. Lett.}
{\bf 49}, 1110 (1982); A.A.~Starobinskii, {\it Phys. Lett. B}
{\bf 117}, 175 (1982); S.W.~Hawking, {\it ibid} {\bf 115}, 295 (1982).

\bibitem{baryo} See e.g., E.W.~Kolb and M.S.~Turner,
{\it Ann. Rev. Nucl. Part. Sci.} {\bf 33}, 645 (1983), or
A.~Cohen, D.~Kaplan, and A.~Nelson, {\it ibid} {\bf 43},
in press (1993).

\bibitem{gamow} G.~Gamow, {\it Phys. Rev.} {\bf 70}, 527
(1946); R.~Alpher, H.~Bethe, and G.~Gamow, {\it ibid}
{\bf 73}, 803 (1948); Ya.B.~Zel'dovich, {\it Sov. Phys. Usp.}
{\bf 6}, 475 (1963).  Zel'dovich's paper is interesting because
not only did he predict the existence of the CBR on the basis
of primordial nucleosynthesis, but misinterpreting observational
upper limits to a background temperature he rejected the hot
big-bang model.

\bibitem{walkeretal} T.P.~Walker et al., {\it Astrophys. J.}
{\bf 376}, 51 (1991).

\bibitem{mass} See e.g., V.~Trimble, {\it Ann. Rev. Astron.
Astrophys.} {\bf 25}, 425 (1987); K.~Ashman, {\it Proc. Astron.
Soc. Pac.} {\bf 104}, 1109 (1992).

\bibitem{rubin} See e.g., V.~Rubin, {\it Proc. Natl. Acad. Sci.
USA} {\bf 90}, 4814 (1993).

\bibitem{iau} See e.g., {\it Dark Matter in the Universe},
eds. J.~Kormendy and G.R.~Knapp (Reidel, Dordrecht, 1985).

\bibitem{dipole} A.~Kogut et al., {\it Astrophys. J.},
in press (1993).  Earlier measurements of the dipole anisotropy
include G.~Smoot, M.~Gorenstein, and R.~Muller, {\it Phys. Rev.
Lett.} {\bf 39}, 14 (1977); D.J.~Fixsen, E.S.~Cheng, and D.T.~Wilkinson,
{\it ibid} {\bf 50}, 620 (1983); R.~Fabbri et al., {\it ibid} {\bf 44},
1563 (1980); A.A.~Klypin et al., {\it Sov. Astron. Lett.} {\bf 13},
104 (1987).

\bibitem{iras}  N.~Kaiser et al., {\it Mon. Not. Roy. astr. Soc.}
{\bf 252}, 1 (1991);
M.~Strauss et al., {\it Astrophys. J.} {\bf 385}, 444 (1992).

\bibitem{relic} See e.g., M.S.~Turner, {\it Physica Scripta}
{\bf T36}, 167 (1991).

\bibitem{seesaw} M.~Gell-Mann, P.~Ramond, and R.~Slansky,
in {\it Supergravity}, eds. D.~Freedman and P.~van Nieuwenhuizen
(North-Holland, Amsterdam, 1979); T.~Yanagida,
in {\it Proceedings of the Workshop on Unified Theories
and Baryon Number in the Universe}, eds. O.~Sawada and
A.~Sugamoto (KEK, Tsukuba, Japan, 1979).

\bibitem{susy} H.E.~Haber and G.L.~Kane, {\it Phys. Rep.}
{\bf 117}, 75 (1985); J.~Ellis et al., {\it Nucl. Phys. B}
{\bf 238}, 453 (1984).

\bibitem{PQ} R.D.~Peccei and H.R.~Quinn, {\it Phys. Rev.
Lett.} {\bf 38}, 1440 (1978); S.~Weinberg, {\it ibid} {\bf 40}, 223
(1978); F.~Wilczek, {\it ibid} {\bf 40}, 279 (1978);
J.-E.~Kim, {\it ibid} {\bf 43}, 103 (1979).

\bibitem{axion1} J.~Preskill, M.~Wise, and F.~Wilczek,
{\it Phys. Lett. B} {\bf 120}, 127 (1983); L.~Abbott and P.~Sikivie,
{\it ibid} {\bf 120}, 133 (1983); M.~Dine and W.~Fischler,
{\it ibid} {\bf 120}, 137 (1983).

\bibitem{axion2}  M.S.~Turner, {\it Phys. Rep.} {\bf 197},
67 (1990).

\bibitem{dmsearch} J.R.~Primack, D.~Seckel, and B.~Sadoulet,
{\it Ann. Rev. Nucl. Part. Sci.} {\bf 38}, 751 (1988);
P.F.~Smith and J.D.~Lewin, {\it Phys. Rep.} {\bf 187}, 203 (1990).

\bibitem{halodm}  D.O.~Caldwell, {\it Mod. Phys. Lett. A} {\bf 5}, 1543 (1990);
K.~van Bibber et al., in {\it Trends in Astroparticle
Physics}, eds. Cline, D. \& Peccei, R.D. (World Scientific, Singapore, 1992),
p.~154.

\bibitem{numass}  See e.g., F.~Boehm and P.~Vogel, {\it Ann. Rev.
Nucl. Part. Sci.} {\bf 34}, 125 (1984).

\bibitem{PIB}  P.J.E.~Peebles, {\it Nature} {\bf 327}, 210 (1987);
{\it Astrophys. J.} {\bf 315}, L73 (1987); R.~Cen, J.P.~Ostriker,
and P.J.E.~Peebles, {\it ibid}, in press (1993).

\bibitem{hdm} S.D.M.~White, C.~Frenk, and M.~Davis,
{\it Astrophys. J.} {\bf 274}, L1 (1983); {\it ibid}
{\bf 287}, 1 (1983); J.~Centrella and A.~Melott, {\it Nature}
{\bf 305}, 196 (1982).

\bibitem{cdm} See e.g., G.R.~Blumenthal et al., {\it Nature}
{\bf 311}, 517 (1984); J.P.~Ostriker, {\it Ann. Rev. Astron. Astrophys.}
{\bf 31}, in press (1993).

\bibitem{MDM}  Q.~Shafi and F.~Stecker,
{\it Phys. Rev. Lett.} {\bf 53}, 1292 (1984);
A.~van Dalen and R.K.~Schaefer, {\it Astrophys. J.}
{\bf 398}, 33 (1992); M.~Davis, F.~Summers, and D.~Schlegel,
{\it Nature} {\bf 359}, 393 (1992); A.~Klypin et al.,
{\it Astrophys. J.}, in press (1993).

\bibitem{top}  A.~Albrecht and A.~Stebbins, {\it Phys.
Rev. Lett.} {\bf 69}, 2615 (1992); D.~Bennett, A.~Stebbins,
and F.~Bouchet, {\it Astrophys. J.} {\bf 399}, L5 (1992);
N.~Turok, {\it Phys. Rev. Lett.} {\bf 63}, 2652 (1989);
A.~Gooding, D.~Spergel, and N.~Turok,
{\it Astrophys. J.} {\bf 372}, L5 (1991).

\end{thebibliography}
\end{document}